\def\bea{\begin{eqnarray}}
\def\eea{\end{eqnarray}}
\def\be{\begin{equation}}
\def\ee{\end{equation}}
\def\bea{\begin{eqnarray}}
\def\eea{\end{eqnarray}}
\def\d{\delta}
\def\p{\phi}
\def\s{\rm sech}
\def\t{\rm tanh}
\documentclass[12pt]{article}
\title { \bf Classical Solution for the Bounce Up to Second Order}
\author{Hatem Widyan \thanks{E--mail : widyan@ahu.edu.jo } and {Mashhoor Al-Wardat} \\
    {Department of Physics} \\
    {Al-Hussein Bin Talal University} \\
     {P.O.Box 20, 71111, Ma'an, Jordan}
    }

\begin{document}
\maketitle
%
%
%
{\bf Abstract:} Scalar field theory with asymmetric potential is
studied  for $\phi^4$ theory with $\phi^3$ symmetry breaking. The
equations of motion are  solved analytically up to the second
order to get the bounce-solution.



{\bf keywords:} phase transition, tunnelling, scalar field theory
%
%
\begin{section}{\bf Introduction}

Vacuum decay is an old subject in field theory \cite{Langer}.
Coleman and Callan \cite{Coleman} showed that a quantum tunneling
process from a false vacuum to a true vacuum can be realized via
the nucleation of a true vacuum bubble in the surrounding of a
false one. Coleman and De Luccia \cite{{Luccia}}  found that
gravity has a significant effect on the vacuum decay process.

In semiclassical approximation, the decay rate per unit volume is
given by an expression of the form
\begin{equation}
  \Gamma \, = \, A \, {\rm e}^{-S_E/\hbar}[1+O(\hbar)] , \label{eq:decay}
\end{equation}
where $S_E$ is the Euclidean action for the bounce: the classical
solution of the equation of motion with appropriate boundary
conditions and the prefactor $A$, comes from Gaussian functional
integration over small fluctuations around the bounce, and has
been discussed in Ref. \cite{Luccia, Weinberg,Munster}. The
solution of the equation of motion looks like a bubble in four
dimensional Euclidean space with radius $R$ and, thickness
proportional to the coefficient of the symmetry breaking term in
the potential. If there is more than one solution satisfying the
boundary conditions, then that with the lowest $S_E$ dominates
Eq.~(\ref{eq:decay}).

Recently, some authors have discussed the vacuum decay in
different situations such as: different scalar field theories
\cite {Hatem}, nonminimal coupling between the scalar field and
curvature scalar \cite{Lee}, DBI action \cite{Brown}, and
non-thin-wall limit \cite{Gen}, etc. The finite temperature effect
on the false vacuum decay process has also been discussed by Linde
et. al. \cite{Linde}, where one should look for the $O
(3)$-symmetric solution due to periodicity in the time direction
with the inverse temperature period $T^{-1}$, instead of the $O
(4)$-symmetric solution at zero temperature. The cosmological
applications of false vacuum decay process have been applied to
various inflation cosmological models \cite{Guth}.

In general, it is not possible to find an analytical solution of
the field equations for a finite difference of the potential
between two minima. An approximation, where the field equations
can be solved exactly, is the "thin wall approximation" (TWA)
\cite{Coleman}. An analytical calculation of the nucleation rate
for
 first order phase transitions beyond the TWA of the standard
 Ginzburg-Landau potential with $\phi$ asymmetric term
 were studied by M$\ddot{\rm u}$nster and Rotsch \cite{Munster}.
 In this paper, we extend our earlier work  for the $\phi^4$ theory
with $\phi^3$ symmetry breaking term \cite{Hatem}, and we
calculate the bounce and the radius of the bubble up to second
order by expanding the bubble solution in powers of the asymmetry
parameter.

The paper is organized as follows. In Sec. 2 we present the
Euclidean action and the equation of motion of the scalar field
$\phi$. In Sec. 3  and 4 we calculate the bounce and radius of the
bubble as well as the action. Sec. 5 includes our conclusion and
discussion.

\end{section}
%
%
\begin{section}{\bf Equation of the motion and the action }

Let us consider a three-dimensional scalar field theory with a
Lagrangian density

\be
 \mathcal{L}=\frac{1}{2} \partial_\mu \phi \partial_\mu \phi-
 U_s(\phi), \label{lagrange}
\ee
 where $U_s(\phi)$ is a symmetric double-well potential having
two degenerate minima at $\phi=\pm 1$ and has the from

\be
 U_s(\phi)= \frac{1}{2} (\phi^2 -1)^2.
\ee
To shift mutually the minima, we introduce a small asymmetry  term
proportional to $\delta$:

\begin{equation}
 U(\phi) = {1 \over 2}( \phi^2 - 1)^2 +
 \delta \phi^3 , \label{potential}
\end{equation}
where $0 \le \delta \le 2.$ The parameter $\delta$ fixes the
asymmetry of the potential. In particular, the difference between
the values of the potential $\phi_\pm$ is
\be
 U(\p_+)-U(\p_-)=2 \d + O(\d^3)
\ee
The shift of the minima in terms of the asymmetry parameter $\d$
is given by
\be
 \p_\pm=\pm1-\frac{3}{4} \d \pm \frac{9}{32} \d^2+O(\d^4) \label{shift}
\ee
%
The field equation for a radially symmetric field is
\begin{equation}
-{d^2\phi \over dr^2}-{2 \over r} {d\phi \over dr}+2 \phi^3-2
\phi^2 + 3 \delta \phi^2=0, \label{eom}
\end{equation}
with the boundary conditions

\be
 \p \rightarrow \p_+ \quad {\rm as} \quad r \rightarrow \infty,
 \quad \frac{d\p}{dr}=0 \quad {\rm
 at} \quad r=0. \label{bc}
\ee
Using the expression  of $\p$ we can calculate the Euclidean
action of the bounce, which is given by
\bea
 S_E(\p)&=& \int d^3 [{1 \over 2} \partial_\mu\p \partial_\mu\p +
 U(\p)] \nonumber \\
 &=& 4\pi \int_0^\infty dr \, r^2 \, \Big({{1 \over 2}({d\phi \over dr})^2+
 [{1 \over 2}(\p^2-1)^2-{1 \over 2}(\p_+-1)^2]+\d (\p^3
 -\p_+^3)}\Big). \label{action}
\eea
\end{section}
%
%
\begin{section}{\bf The bounce solution}

Before applying the systematic approach to solve Eq.~(\ref{eom}),
let us consider the thin wall approximation. This usually happens
when $\d$ is small, and gives a solution  nearly equals to $-1$
inside a sphere of radius $R$ and nearly equal to $+1$ outside it.
The region where $\p$ differs significantly from these values is a
kink of the from
\be
 \p(r)= {\rm tanh}(r-R), \label{kink}
\ee
which is the solution of the field equation

\begin{equation}
-{d^2\phi \over dr^2}-{2 \over r} {d\phi \over dr}+2 \phi^3-2
\phi^2=0.
\end{equation}

One can easily show  that the energy of the bubble (the action of
the bounce solution) of radius $R$ in the thin wall approximation
(see for example \cite{Coleman}) is given by

\be
 S_E(R)=-\frac{4}{3} \pi R^3 \epsilon+ 4 \pi R^2 S_1,
\ee
where $\epsilon=2f$ and $S_1$ is the bubble-wall surface energy
(surface tension), and is given by
\be
 S_1= \int dr \, \{{1 \over 2}({d\phi \over dr})^2+U_s(\p)\}.
\ee
The critical bubble radius $R$, for which the energy is
stationary, is written in terms of $S_1$ and $\epsilon$ as
\be
 R=\frac{2 S_1}{\epsilon}=\frac{4}{3 }{1 \over \d}, \label{R0}
\ee
whence it follows that
\be
 S_E=\frac{16 \pi S_1^3}{3 \epsilon^2}=\frac{256
 \pi}{81}\frac{1}{\d^2}. \label{S0}
\ee

Note that for in the thin wall approximation (i.e. small
$\delta$), we have exact analytical result for the radius as well
as for the action. The radius is large and therefore the wall is
indeed thin compared to the size of the bubble.

For finite $\delta$, the solution of the equation of motion
Eq.~(\ref{eom}) can't be written in a closed from. Following the
approach in \cite{Munster}, the solution is constructed by means
of an expansion in powers of $\delta.$

Since the bounce-solution is centered around the radius $R$, then
it is more convenient to write the equation of motion in terms
$\xi=r-R$. Hence Eq.~(\ref{eom}) becomes

\begin{equation}
-{d^2\phi \over d\xi^2}-{2 \over {\xi+R}} {d\phi \over d\xi}+ 2
\phi^3-2 \phi + 3 \delta \phi^2=0
\end{equation}

Depending on the thin wall approximation, we may write a Laurent
series as an ansatz for the critical radius,

\begin{equation}
R={a_{-1} \over \delta} + a_0+ a_1 \delta + a_2 \delta^2+...
=\sum_{i=-1}^\infty a_i \d^i. \label{radius}
\end{equation}

The factor of the first derivative in the equation of motion
becomes
\be
 \frac{1}{\xi+R}=\frac{1}{\xi+\sum_{i=-1}^\infty a_i
 \d^i}=\frac{\d}{a_{-1}}-\frac{\xi+a_0}{a^2_{-1}}\d^2+
 \frac{(\xi+a_0)^2-a_1 a_{-1}}{a^3_{-1}}
 \d^3+O(\d^4).
\ee

 Hence the expansion of the field equation in powers of $\delta$ reads

\begin{equation}
-{d^2\phi \over d\xi^2}-{2 \over a_{-1}}\delta {d\phi \over d\xi}+
 {2(\xi+a_0) \over a_{-1}^2}\delta^2 {d\phi \over d\xi}+2\phi^3-2
\phi + 3\delta \phi^2+ O(\delta^3)=0.
\end{equation}

Now, the solution of the above equation is obtained perturbatively
up to the second order by means of the expansion

\begin{equation}
\phi=\phi_0+\delta \phi_1 + \delta^2 \phi_2 + O(\delta^3).
\end{equation}

For the zero order of $\d$ the equation of motion is

\be
 -{d^2\phi_0 \over d\xi^2}+2 \p_0^3-2 \p_0=0,
\ee
 which has  kink's solution
 \be
  \p_0(\xi)= {\rm tanh}\xi.
 \ee
Note that the solution satisfies the boundary conditions given in
Eq.~(\ref{bc}), i.e.,   $\p_0(\infty)=+1$ and $\p'_0(0)=0.$

The equation of motion for the first order of $\d$
\be
  -{d^2\phi_1 \over d\xi^2}-(6\, \s^2\xi-4)\p_1=({2 \over
  a_{-1}}+3) \,\s^2\xi-3. \label{eom1}
\ee
The homogenous part of Eq.~(\ref{eom1}) has the following two
independent solutions
\bea
 \p_{1,H_1}(\xi)&=&-C_1\, \s^2\xi \nonumber
\\
 \p_{1,H_2}(\xi)&=&C_2 (\frac{3}{2}\xi\, \s^2\xi+{3 \over 2} \t\xi +{\rm
 sinh}\xi \, {\rm cosh}\xi). \nonumber
\eea
The particular solution of the equation which satisfies the
boundary conditions for $\xi \rightarrow \infty$ is
$$
 \p_{1,S}=C_3+C_4 \,\s^2\xi
 $$
 Substituting $ \p_{1,S}$ in Eq.~(\ref{eom1}), we get
 \be
  a_{-1}={4 \over 3},
\ee
 which fixes the leading coefficient in $R$, Eq.~(\ref{radius}),
 and
\be
  \p_{1,S}= -{3 \over 4}.
\ee
The general bounce-solution is given in the first order of $\d$ by
\be
 \p_1(\xi)=  - C_1\, \s^2\xi+ C_2 (\frac{3}{2}\xi\, \s^2\xi+{3 \over 2}
  \t\xi +{\rm sinh}\xi \, {\rm cosh}\xi) -{3 \over 4}.
\ee
Note that the constant term reflects the shift of the minimum
given in Eq.~(\ref{shift}). The term proportional to $C_2$
diverges for large $\xi$, hence $C_2$ must be zero. The first term
corresponds to the derivative of the zero-order of the
bounce-solution, and is related its translation degree of freedom.
To see this, we expand the zero-order solution around $\d=0$ in a
Taylor series as
\be
 \t(\xi-C_1 \d)=\t\xi- C_1\,\d\,\s^2\xi - C_1^2\, \d^2\, \s^3\xi \,{\rm
 sinh}\xi +O(\d^3).
\ee
 Note that $C_1$ corresponds to the parameter $a_1$ in the ansatz
 for the radius Eq.~(\ref{radius}). The homogenous solution proportional
 to  $C_1$ is already taken into consideration by this ansatz, and
 should not consider in the the following calculations. Hence, we remain
 with
\be
   \p_1(\xi)=  -{3 \over 4}.
\ee
Note that $\p_1$ satisfies the boundary conditions given in
Eq.~(\ref{bc}).

The equation of motion for the second order of $\d$
\be
  -{d^2\phi_2 \over d\xi^2}+(6\, \s^2\xi-2)\p_2=-{9 \over
  8} \, \xi\, \s^2\xi-{3 \over 2}\, a_0\, \s^2\xi + {9 \over 8} \,\t\xi,
  \label{eom2}
\ee
which has a general solution
\bea
  \p_2(\xi) &=& - D_1\, \s^2\xi+ D_2 (\frac{3}{2} \, \xi \, \s^2\xi+{3 \over 2}\, \t\xi +
  {\rm sinh}\xi \, {\rm cosh}\xi)+ \xi ({3 \over 16}+ {3\over 16}\, {\rm cosh}^2\xi-{15\over 64 }
   \,\s^2\xi ) \nonumber \\
 & - & {\rm Log}\,{\rm cosh}\xi ({9 \over 32} \,\xi \, \s^2\xi + {9 \over 32}\, \t\xi +
 {3 \over 16} \,{\rm sinh}\xi \, {\rm
 cosh}\xi) + {3 \over 64} \, \t\xi - {3 \over 16}\, {\rm sinh\xi}\,
 {\rm cosh}\xi \nonumber \\
 & +& a_0({9 \over 2} + 4 \, {\rm sinh}^2\xi \, \t^2\xi) +{9 \over
 32} \, \s^2\xi \, {\it T(\xi)},
\eea
where we define
\be
 T(\xi) = \int_0^\xi \xi' \, \t\xi' \, d\xi'.
\ee
The general solution has a converging terms as well as diverging
terms for large values of $\xi$. As the same argument for $\p_1$,
$D_1=0$. The term proportional to $a_0=0$ is symmetric in $\xi$,
while the diverging homogenous solution of the bounce is
antisymmetric, so we set $a_0=0$ and we assume $D_2={3 \over
16}-{3 \over 16}\, {\rm Log}2$ in order to get rid of the
diverging terms. Therefore the converging solution is given by
\bea
 \p_2(\xi) &=& - {9 \over 32} \, \xi (\t\xi-1) + {3\over 32 }\,
 \xi ( {\rm cosh}\xi -{\rm sinh}\xi )^2-{9 \over 32}\, \xi^2 \,
 \s^2\xi + { 3 \over 64} \, \xi \, \s^2\xi + {21 \over 64} \, \t\xi \nonumber \\
 & - & {\rm Log}(1+{\rm e}^{-2\xi} ) ({9 \over 32} \,\xi \, \s^2\xi
  + {9 \over 32}\, \t\xi+ {3 \over 16} \,{\rm sinh}\xi \, {\rm
 cosh}\xi    ) +{9 \over
 32} \, \s^2\xi \, {\it T(\xi)}.
\eea
In order check whether $\p_2$ is an acceptable solution or not, we
find its values at a large values of $\xi$, and its derivative at
$\xi=-R$. It can be easily shown that
\be
 \p_2 (\xi \rightarrow \infty) = {9 \over 32} ,
\ee
which reflects the shift of the minimum given in
Eq.~(\ref{shift}), and
\be
 \p_2'(\xi=-R)=0.
\ee
Hence $\p_2$ satisfies the boundary conditions given in
Eq.~(\ref{bc}), which means that it is an acceptable solution.

Whereas the first order solution corresponds to the shift of the
minimum and the critical radius, the second order solution
describes a true deformations of the bubble. The boundary
condition at $r=0$, i.e. $\xi=-R$, is fulfilled order by order in
$\d$. For example, the leading order solution yields

\be
 \p'_0(-R)={\rm e}^{-2/\d}(4+O(\d)),
\ee
 which vanishes to all orders in $\d$. Similar observations hold
 in higher orders.

The bounce solution up to the second order of $\d$ is
\bea
 \p(\xi)&=&\phi_0+\delta \phi_1 + \delta^2 \phi_2 + O(\delta^3).
 \nonumber \\
   &=& {\rm tan}\xi- \d {3 \over 4} +\d^2 \Big(- {9 \over 32} \, \xi (\t\xi-1) + {3\over 32 }\,
 \xi ( {\rm cosh}\xi -{\rm sinh}\xi )^2-{9 \over 32}\, \xi^2 \,
 \s^2\xi + { 3 \over 64} \, \xi \, \s^2\xi  \nonumber \\ &+& {21 \over 64} \, \t\xi
  -  {\rm Log}(1+{\rm e}^{-2\xi} ) ({9 \over 32} \,\xi \, \s^2\xi
  + {9 \over 32}\, \t\xi+ {3 \over 16} \,{\rm sinh}\xi \, {\rm
 cosh}\xi    ) \nonumber \\ &+&{9 \over
 32} \, \s^2\xi \, {\it T(\xi)}\Big) + O(\delta^3).
\eea

\end{section}

%
\begin{section}{\bf The action and the critical radius}

With the expression for $\p$ we can calculate the action of the
bounce, which is given by Eq.~(\ref{action}) as
 \bea
 S_E(\p) &=& 4\pi \int_0^\infty dr \, r^2 \, \Big({{1 \over 2} \p'^2+
 \Big[{1 \over 2}(\p^2-1)^2-{1 \over 2}(\p_+-1)^2\Big]+\d (\p^3  -\p_+^3)}\Big)
 \nonumber \\
 &=& S_0 + \d S_1 + \d^2 S_2 + O(\d^3).
\eea
The integrands are centered around the critical radius $r=R$. The
integration range in $\xi$ can be extended to the whole real axis
\cite{Munster}. Hence
\bea
 S_0 &= &4\pi \int_{-\infty}^\infty d\xi \, (\xi+R)^2 \, \Big({1 \over 2}\p_0'^2+
 \Big[{1 \over 2}(\p_0^2-1)^2-{1 \over 2}(\p_{0+}-1)^2\Big]\Big)
 \nonumber \\
  &=& 2 \pi ({8 \over 3} R^2 - {{12-2\pi^2} \over 9}).
\eea

\bea
 S_1 &= &4\pi \int_{-\infty}^\infty d\xi \, (\xi+R)^2 \, \Big(\p_0'
 \p_1'+2 \p_0 \p_1 (\p_0^2-1) + (\p_0^3-1) \Big) \nonumber \\
  &=& 2 \pi (-{4 \over 3} R^3 +  (2-{\pi^2 \over 3})R).
\eea

\bea
 S_2 &= &2\pi \int_{-\infty}^\infty d\xi \, (\xi+R)^2 \, \Big(
 \p_1'^2+ 2 \, \p_0' \, \p_2'+ 6 \, \p_0^2 \, \p_1^2 - 2 \, \p_1^2
 + 4 \, \p_0 \p_2 (\p_2^2-1)-{9 \over 4} + {9 \over 2}
 (1-\p_0^2)\Big)
  \nonumber \\
  &=& 2 \pi (-{9 \over 4} R^2 + 1.01528).
\eea
So, all together, the action is given up to the second order of
$\d$ as
\be
 S_E= 2 \pi \Big[{8 \over 3} R^2 - {{12-2\pi^2} \over 9} + \d \Big(
  -{4 \over 3} R^3 +  (2-{\pi^2 \over 3})R \Big)
 +\d^2 \Big(-{9 \over 4} R^2 + 1.01528 \Big) \Big].
\ee

To find an expression for the critical radius $R$, we minimize the
action with respect to $R$, i.e.,
\be
 {dS_E \over dR}=0.
\ee
Explicit calculation of the coefficients leads to two more terms
in the Laurent series in $R$:
\be
 R={ 4 \over 3} {1 \over \d}+0+ \d ({11 \over 64}-{\pi^2 \over
 16})+ 0 . \d^2 + O(\d^3), \label{R}
\ee
so that the bubble is now completely determined to the second
order. Hence, the action follows as
\be
 S_E=2 \pi \Big[ {128 \over 81} { 1 \over \d^2}- ({8 \over 3} +
  {2 \pi^2 \over 9}) + O(\d^2) \Big]. \label{S} \ee

For the thin wall approximation, we have obtained an explicit
values of the radius and  the bounce which are given in
Eq.~(\ref{R0}) and Eq.~(\ref{S0}) respectively, these are exactly
the values of the leading terms in Eq.~(\ref{R}) and
Eq.~(\ref{S}).

To check the validity of our analytical results, we have compared
our analytical results with the thin-wall approximation for
different values of $\delta$. Figure $1$ shows a plot of
$S_E/S_{TWA}$, we can see that as far as $\delta<0.1$, the ratio
is of order $1$, and for $\delta > 0.1$, the ratio is less than
$1$. The result is physically expected, since the maximum value of
the action is in the case of thin-wall approximation. Similar
observation also for the radius as shown in figure $2$. We have
also solved the equation of motion (Eq.~(\ref{eom})) numerically.
Figure $3$ shows the ratio Euclidean action of the numerical
results to the analytical results. We notice from the figure that
as far as $\delta<0.1$, there is a good agreement between them,
and for $\delta>0.1$, our analytical results starts deviates from
the numerical ones.

\end{section}

%
\begin{section}{\bf Conclusion}

By expanding all   of the quantities in power of $\d$, we found
the bounce solution, the critical radius and the action
analytically beyond the thin wall approximation. We also showed
that the leading terms correspond to the thin wall approximation.

The  comparison between our analytical results and the numerical
ones showed a good correlation up to $\delta=0.1$, and after that
the analytical results starts deviates form the numerical ones.
Also our analytical results are consistent with the TWA results up
to $\delta=0.1$.

 To calculate the nucleation rate, we have to find the fluctuation
 determinant in powers of $\d$, this will be determined in future
 work.

\end{section}

\bibliography{plain}

\begin{figure}[ht]
\vskip 15truecm

\includegraphics{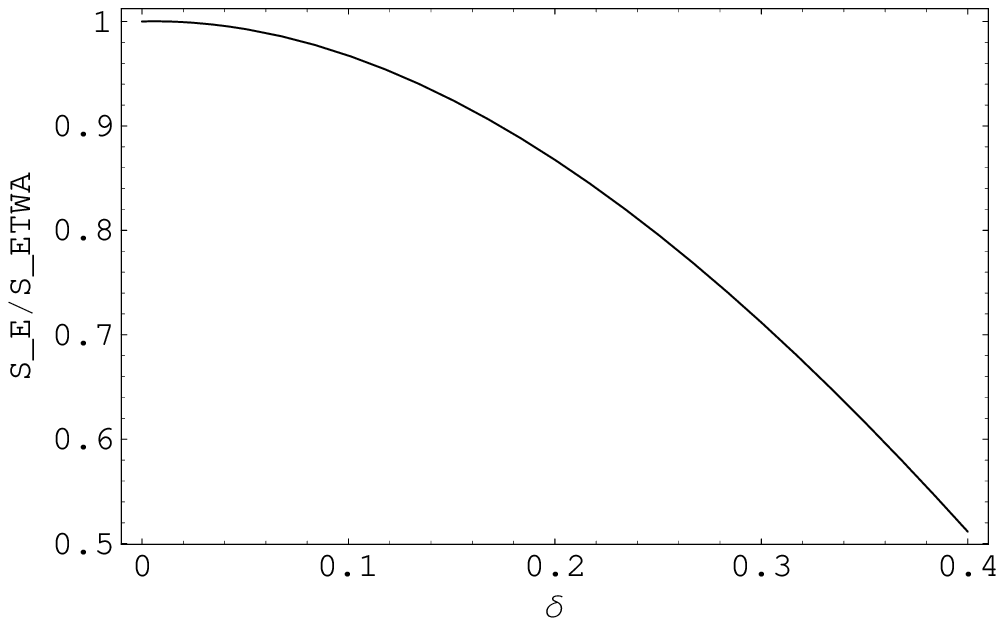}

\caption{Ratio of the analytical action  to the action in the
thin-wall approximation: $S_E/S_{ETWA}$ vs $\delta$}
\end{figure}

\begin{figure}[ht]
\vskip 15truecm

\includegraphics{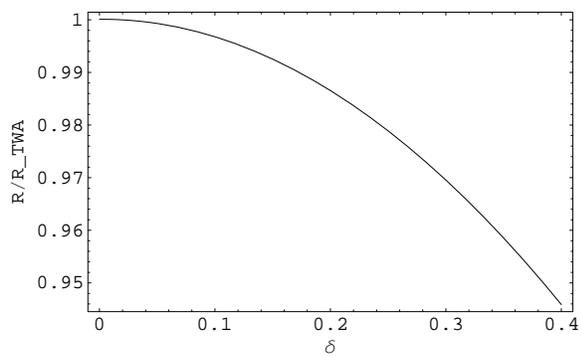}

\caption{Ratio of the analytical radius of the bubble to the
radius in the thin-wall approximation: $R/R_{TWA}$ vs $\delta$}
\end{figure}

\begin{figure}[ht]
\vskip 15truecm

\includegraphics{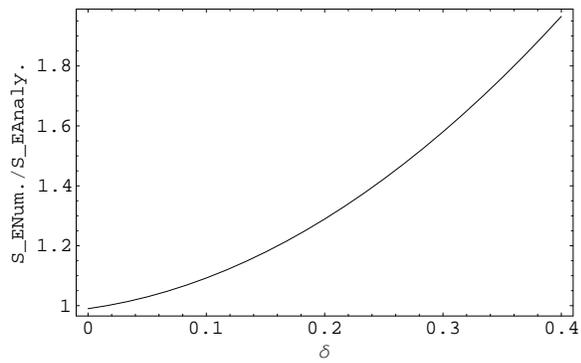}

\caption{Ratio of the numerical action to the analytical action:
$S_{ENum.}/S_{EAnaly.}$ vs $\delta$}
\end{figure}

\end{document}